\documentstyle[12pt]{article}
\newcommand{\fr}{\frac}
\newcommand{\lb}{\label}
\newcommand{\ti}{\tilde}
\newcommand{\be}{\begin{equation}}
\newcommand{\ee}{\end{equation}}
\newcommand{\beqa}{\begin{eqnarray}}
\newcommand{\la}{\lambda}
\newcommand{\del}{\partial}
\newcommand{\eeqa}{\end{eqnarray}}
\newcommand{\ep}{\epsilon}
\newcommand{\p}{{\cal P}}
\newcommand{\Omo}{\Omega}

\begin{document}
\title{BV and BFV Formulation of a Gauge Theory of Quadratic Lie
Algebras in 2-d and a Construction of $W_3$ Topological Gravity}
\author{\"{O}mer F. DAYI \\
Physics Department, T\"{U}B\.{I}TAK Marmara Research Centre,\\
P.O.Box 21, 41470 Gebze, Turkey\thanks{E-mail address:DAYI@TRMBEAM.bitnet}}
\date{}
\maketitle

\begin{abstract}

The recently proposed generalized field method for
solving the master equation of Batalin and Vilkovisky is applied
to a gauge theory of quadratic Lie algebras in 2-dimensions.
The charge corresponding to BRST symmetry
derived from this solution
in terms of the phase space variables by using the Noether procedure,
and the one found due to the BFV-method are compared and found to
coincide. $W_3$ algebra, formulated in terms of a continuous
variable is emploied in the mentioned
gauge theory to construct a $W_3$ topological
gravity. Moreover, its gauge fixing  is briefly discussed.

\end{abstract}

\pagebreak

\section{Introduction}

Gauge theries of Lie groups offered an understanding of elementary particles
physics. Going beyond Lie groups can be obtained by deforming their
algebra. The simplest deformation is to let the commutator of two
generators possess some quadratic terms in the generators,
in addition to the linear ones.
Obviously this is not a Lie algebra,
nevertheless, it is known as ``quadratic Lie algebra".
Yang-Mills theory is not suitable to formulate a gauge theory
of quadratic Lie algebras.
A gauge theory Lagrangian of quadratic Lie algebras is proposed in
 \cite{II}, by utilizing the formulation given in  \cite{SSN}.
Its gauge algebra is closed on mass shell.

The most powerful methods of quantization of the gauge systems
possessing a gauge algebra closing
on mass shell and/or irreducible gauge generators
are  Batalin-Vilkovisky (BV)\cite{BV} and Batalin-Fradkin-Vilkovisky
(BFV)\cite{BFV} schemes. The former is a Lagrangian and the latter is
a Hamiltonain formulation.

The first step in the BV-method is to find the proper solution of the master
equation. Usually this is a hard task. Fortunately, for a vast class of
first order systems a general solution is known\cite{O12}. We apply
this method to the gauge theory of quadratic Lie algebras and find
the proper solution of the master equation. By using this solution
and the Noether procedure we find a charge corresponding
to BRST symmetry, in terms of the related phase space variables.
On the other hand we can
find a BRST charge of this system following from the BFV-method.
{\it A priori} there is not any argument to conlude that these
two charges are
the same\footnote{The relation between BV- and BFV- method
is studied in  \cite{rel}. They give the solution of the
master equation if the BFV-BRST charge is known.}.
Here, we show that these two charges coincide for the gauge
theory of quadratic Lie algebras.

The gauge theory Lagrangian which we deal with does not depend
on the space-time metric and does not lead to local excitations.
Hence it is a topological quantum field theory
(for a review see  \cite{BBRT}).

$W_3$ algebra\cite{Z} is an example of quadratic Lie algebras.
Gauging this algebra has received a lot of attention\cite{gW3}.
In  \cite{SSN} beginning from the classical $W_3$
algebra, which possesses infinite  generators\footnote{Gauging
Lie algebras which have infinite generators are studied in \cite{il}.},
a gauge theory is built without refering a Lagrangian.
As a bookkeping device they introduced a  continuous variable,
so that they could write the gauge transformations in a
compact form. They also pointed out that a topological $W_3$ gravity
based on their formulation is expected to be possible.
Here, we give a formulation of $W_3$ topological gravity in this spirit.
But, we employ the continuos variables from the beginning to
write $W_3$ algebra commutators, instead of introducing them as a
bookkeping device. Obviously our construction is different from
the topological $W_3$ gravity formulation based
on $sl(3)$  given in \cite{Li}.

\section{BV and BFV Formulation of a Gauge {Theory} of Quadratic
Lie Algebras in 2-d}

The simplest deformation of a Lie algebra generated by $T_a$ is
\be
[T_a, T_b]={f_{ab}}^c T_c +V_{ab}^{cd}T_cT_d +k_{ab},
\ee
which is  known as quadratic Lie algebra. Here, $f$, $V$, and $k$
are constants and their symmetry properties are
\be
\lb{sym}
{f_{ab}}^c=-{f_{ba}}^c,\   V_{ab}^{cd}=-V_{ba}^{cd},\
V_{ab}^{cd}=V_{ab}^{dc}, \  k_{ab}=-k_{ba}.
\ee
Commutators should satisfy the Jacobi identities, hence the
constants $f$, $V$, and $k$ should be chosen such that
\beqa
{f_{[ab}}^d{f_{c]d}}^e & =0, & \nonumber \\
{f_{[ab}}^dV_{c]d}^{ef} & +V_{[ab}^{df}{f_{c]d}}^e &+
V_{[ab}^{ed}{f_{c]d}}^f =0,  \nonumber \\
V_{[ab}^{de}V_{c]d}^{fg} & =0, &  \lb{ji} \\
{f_{[ab}}^d k_{c]d} & =0, &  \nonumber \\
V_{[ab}^{de}k_{c]d} & =0, &  \nonumber
\eeqa
are obeied. Here, $[\ ]$ denote antisymmetrization in the
indices which are within them.

Gauge theory of this algebra in 2--d space-time is given
by the Lagrange density\cite{II}
\be
\lb{l0}
{\cal L}=-\fr{1}{2}\ep^{\mu \nu}\{ \Phi_a(\del_\mu h_\nu^a
-\del_\nu h_\mu^a +{f_{bc}}^a h_\mu^bh_\nu^c +
V_{bc}^{ad} \Phi_dh_\mu^bh_\nu^c)+k_{ab}h^a_\mu h^b_\nu\},
\ee
which is invariant under the gauge transformations
\beqa
\delta h_\mu^a = \del_\mu \la^a +{f_{bc}}^a h_\mu^b \la^c +2
V_{bc}^{ad} \Phi_dh_\mu^b\la^c, \lb{g1} \\
\delta \Phi_a = {f_{ba}}^c\Phi_c \la^b +
V_{ba}^{cd} \Phi_c \Phi_d \la^b +k_{ab}\la^b,  \lb{g2}
\eeqa
whose algebra is closed on mass shell.

To find the proper solution of the (BV-) master equation
whose classical
limit is given by (\ref{l0}), we can apply the generalized field method
of  \cite{O12}, because, the gauge transformations (\ref{g1})-(\ref{g2})
can be written as
\beqa
\delta h_\mu^a= \ep_{\mu \nu}
\fr{\del^2 {\cal L}}{\del \Phi_a \del h_\nu^b}\la^b, \lb{gtt1} \\
\delta \Phi_a= \ep_{\mu \nu}
\fr{\del^2 {\cal L}}{\del h_\mu^a \del h_\nu^b}\la^b.     \lb{gtt2}
\eeqa
In the method given in  \cite{O12} one introduces
the generalized fields
\beqa
\ti{h} = h_{(1,0)} +\eta_{(0,1)}-\Phi^\star_{(2,-1)}, \\
\ti{\Phi}=-h^\star_{(1,-1)}-\eta^\star_{(2,-2)}+\Phi_{(0,0)},
\eeqa
where the first number in the parenthesis is the order of differential
forms and the second is the ghost number.
$\eta^a$ are the ghost fields, and the star denotes the antifields
as well as the Hodge map. As one can observe the total
degree of the
generalized fields $\ti{h},$ $\ti{\Phi},$
which is the sum of ghost number and
order of differential form are 1 and 0.
Now by replacing the fields $h,\ \Phi$ with the generalized
ones $\ti{h},\ \ti{\Phi}$
one can  obtain
the proper solution of the master equation:
\be
S=-\int d^2x\fr{1}{2}\{ \ti{\Phi}_a(d \ti{h}^a
+{f_{bc}}^a \ti{h}^b\ti{h}^c +
V_{bc}^{ad} \ti{\Phi}_d\ti{h}^b\ti{h}^c)+k_{ab}\ti{h}^a\ti{h}^b \}.
\ee
It is the solution of the master equation, because
$S$ is invariant under the gauge
transformation obtained as the generalization of (\ref{gtt1})-(\ref{gtt2}).
In the multiplication of two identical generalized fields
the field which appears first in the generalized field, will be
the first term in the multiplication.
The proper solution of the master equation
$S$, in components is

\pagebreak

\beqa
S= &\int  d^2x \{ {\cal L} & +h^{\star \mu}_a (\del_\mu \eta^a
+{f_{ba}}^c\Phi_c \eta^b +
2V_{bc}^{ad} \Phi_d h_\mu^b\eta^c) \nonumber \\
& & +\Phi^{\star a} ({f_{ba}}^c\Phi_c\eta^b +
V_{ba}^{cd}\Phi_c\Phi_d\eta^b +k_{ba}\eta^b) \nonumber \\
& & +\eta^\star_a (\fr{1}{2}{f_{bc}}^a \eta^b\eta^c +
V_{bc}^{ad}\Phi_d\eta^b\eta^c ) \nonumber \\
& & -\fr{1}{2} \ep_{\mu \nu} V_{bc}^{ad}
h^{\star \mu}_ah^{\star \nu}_d\eta^b\eta^c\} . \lb{gsm}
\eeqa
One can explicitly check that indeed (\ref{gsm}) is the proper solution
of the master equation, by using the symmetries and the identities
which $f$, $V$, and $K$  satisfy (\ref{sym})-(\ref{ji}).

The proper solution of the master equation $S$, is the
full Lagrangian which can be used in the related path integrals after
gauge fixing. Instead of discussing gauge fixing
on general grounds, we prefer to do it after constructing $W_3$
topological gravity.

Let us deal with  Hamiltonian formalism of the theory. Because of
being first order the non-vanishing Poisson brackets can be
read easily from the Lagrange density (\ref{l0}) as
\[
\{\Phi_a ,h_1^b\}=-\delta_a^b .
\]
Now, one can see that $h_0^a$
behave as  Lagrange multipliers and derivations with respect to them
lead to the constraints
\be
\Psi_a =\del_1 \Phi_a +{f_{ac}}^b\Phi_bh^c_1
+V^{cd}_{ab}\Phi_c \Phi_d h^b_1 +k_{ab}h^b_1.
\ee
Canonical Hamiltonian vanishes, so that there is no other constraint.
By making use of the symmetry properties (\ref{sym}), and the
identities (\ref{ji}) of $f$, $V$, and $k$, one finds that
$\Psi_a$ satisfy
\be
\{\Psi_a , \Psi_b\}= ({f_{ab}}^c +2 V_{ab}^{cd}\Phi_d) \Psi_c,
\lb{ca}
\ee
so that they are first class.

To perform the BFV analysis,
let us enlarge the phase space by
introducing the ghost variables $\eta^a$,
and their canonical conjugate $\p_a$ which satisfy
\[
\{\p_a ,\eta^b \}_+ = \delta^b_a,
\]
where $\{A,B\}_+=(\del A/  \del \p_a)\del B /  \del \eta^a +
(\del A/  \del \eta_a)\del B /  \del \p^a.$
$\eta$, and $\p$ are
anticommuting fields, and their ghost numbers are $1$ and$-1$,
respectively.

The BFV--charge $\Omo$, defined to possess ghost number 1, to be
fermionic and to satisfy
\[
\{\Omo ,\Omo \}_+ =0,
\]
by using  (\ref{sym})-(\ref{ji}) can be found as
\be
\lb{omo1}
\Omo = \eta ^a \Psi_a -\fr{1}{2} \eta^a \eta^b \p_c({f_{ab}}^c +
2V^{cd}_{ab} \Phi_d).
\ee

To understand the relation between the BV and the BFV approaches,
we would like to compare $\Omo$ of the latter
with the Noether charge of the former corresponding
to the BRST symmetry given by
$\delta_lS/\delta \ti{\Phi}$. In the
definition of the latter charge one uses the related equations of motion,
but in the former case the charge is not aware of the equations of motion.
Hence, {\it a priori} their relation is not clear.

The Noether charge, in terms of the phase space variables
$(\ti{P},\ \ti{h}),$ is given by
\be
\Omo_N= \ti{P}_A
\fr{\delta_l S}{\delta \ti{\Phi}_A} -K,
\ee
where the generalized momentum is
\be
\lb{cm}
\ti{P} =\fr{\delta_r S}{\delta \dot{\ti{h}}},
\ee
and $K$ is defined to satisfy
\[
\fr{\del K}{\del \ti{P}_A} -\fr{\del (\delta_l S/  \delta \ti{\Phi}_B)}{
\del \ti{P}_A} \ti{P}_B =0.
\]
{}From (\ref{cm}) one finds that the canonical momenta are
\[
P_{h^a_1} =-\Phi_a,\  P_{h^a_0} =0,\  \p_{\eta^a} = h^{\star a}_0.
\]
In terms of these canonical variables one can show that $K$ is
\[
K=V_{cd}^{ab} (-P_{h_1^a}
\p_b \eta^c\eta^d + P_{h_1^a} P_{h_1^a} h^c_1 \eta^d),
\]
which yields
\[
\Omo_N =\Omo.
\]
Thus we can conclude that BV and BFV approaches lead to the same
BRST charge.

One replaces the generalized Poisson brackets $\{,\},\ \{,\}_+$,
with the commutator $[,]$, or anticommutator $[,]_+$,
to achive operator quantization.
Here, this will lead to some complications in the definition of BRST
operator due to operator ordering problems. Nevertheless, let us
suppose that a suitable operator ordering exists such that $\Omo_{op}$,
the operator resembling the BRST charge $\Omo$, is nilpotent.
Then, solution of its cohomology will give the physical states of
the system under study. Unfortunately, there is not any general
procedure to solve this problem.

Recently a general formulation of nonlinear Lie algebras, which
includes also the one which we studied is given\cite{III}.
Also for this formulation the
method of \cite{O12} is suitable for finding the
proper solution of the BV master equation.

\section{$W_3$ Topological Gravity}

$W_3$ algebra can be given in terms of operator product expansion or
equivalently in terms of its modes. Here we will follow yet
another, but an equivalent description.
The algebra is defined in terms of the generators
$G_A(z)=(T(z), W(z) )$ as
\be
\lb{w3b1}
[G_A(z), G_B(w)]=\int  dv
{f_{AB}}^C(z,w,v)G_C(v) +
\int  dv drV^{CD}_{AB}(z,w,v,r)G_C(v)G_D(r) ,
\ee
where ${f_{AB}}^C$, and $V_{AB}^{CD}$ are given as
\beqa
{f_{11}}^1(z,w,v) & = & \del_z\delta (z-w) \delta (z-v) -
\del_w\delta (w-z) \delta (w-v) \lb{f} \\
{f_{12}}^2(z,w,v) & = &3\del_z\delta (z-w) \delta (z-v) +2
      \delta (z-w)\del_z \delta (z-v) \\
V^{11}_{22}(z,w,v,r) & =  &
\delta (z-v) \delta (z-r) \del_z \delta (z-w) \nonumber \\
& &  -\delta (w-v) \delta (w-r) \del_w \delta (z-w). \lb{w3bl}
\eeqa
Obviously we deal with the ``classical" $W_3$ algebra. To obtain
the mode expansion  take
\[
T(z)=\sum_{n=-\infty}^\infty T_n z^{-n-2},\
W(z)=\sum_{n=-\infty}^\infty W_n z^{-n-3},
\]
and perform the integrals (without summation on $A$ and $B$)
\[
\int dz dw\ [G_A(z),G_B(w)] u_A(z) u_B (w),
\]
where
\be
u_1(z)=z^{n+1},\  u_2(z)=z^{n+2}.
\ee
These lead to the known description of classical $W_3$ algebra,
\beqa
{[T_m,T_n]} & = & (n-m) T_{n+m}, \lb{w31} \\
{[T_m,W_n]} & = & (n-2m)W_{n+m}, \\
{[W_m,W_n]} & = & (n-m) \sum_{k=-\infty}^\infty T_{m-k} T_k. \lb{w33}
\eeqa

Although, for the algebra given in (\ref{w31})-(\ref{w33})
Jacobi identities are satisfied, it is not sufficent to
conclude that the algebra defined in (\ref{w3b1})-(\ref{w3bl})
also satisfies Jacobi identities. Because, one can employ
different ${f_{AB}}^C$, and $V_{AB}^{CD}$ which lead to the
same algebra in terms of the modes. This feature is a result of
using continuous variables in the definition of the algebra: we
should suppose that all of the  functions which
take value in the algebra
(depend on the continuous parameters) behave such that
partial integrals are allways allowed.
Nevertheless, after a tedious calculation one can show that
\beqa
\int dr &
[ {f_{11}}^1(z,w,r) {f_{11}}^1(r,v,q) &+
{f_{11}}^1(w,v,r) {f_{11}}^1(r,z,q) \nonumber \\
& &+ {f_{11}}^1(v,z,r) {f_{11}}^1(r,w,q) ] =0,  \nonumber \\
\int dr &
[ {f_{12}}^2(z,w,r) {f_{12}}^2(v,r,q) &+
{f_{21}}^2(w,v,r) {f_{12}}^2(z,r,q)     \nonumber \\
& & + {f_{11}}^1(v,z,r) {f_{21}}^2(w,r,q)] =0,  \nonumber  \\
\int dr & [
V^{11}_{22}(z,w,v,r){f_{21}}^2(p,r,q) &+
 V^{11}_{22}(w,p,v,r){f_{21}}^2(z,r,q)  \nonumber \\
& & + V^{11}_{22}(p,z,v,r){f_{21}}^2(w,r,q) ] =0 .\nonumber
\eeqa
One can also check that $f,$ and $V$ possess the desired symmetry
properties (\ref{sym}).

Now, to formulate a topological $W_3$ gravity let us introduce
\[
h_\mu^A=(e_\mu, B_\mu),\  \Phi_A=(t, w).
\]
Using these fields, and (\ref{f})-(\ref{w3bl}) in (\ref{l0})
yields
\be
S_0= \int d^2x \int dz
\ep^{\mu \nu} (t\del_\mu e_\nu +w\del_\mu B_\nu
+ te_\mu \del e_\nu +3we_\mu\del B_\nu
+2\del w e_\mu B_\nu +tt B_\mu\del B_\nu ),
\ee
where $\del$ is the derivative with respect to $z$.
Thus introduce the generalized fields
\beqa
\ti{h} & = & (e_\mu, B_\mu \oplus \eta_1,\eta_2 \ominus
t^\star,w^\star), \nonumber\\
\ti{\Phi} & = & (\ominus e^{\star \mu},B^{\star \mu} \oplus
\eta^\star_1, \eta^\star_2
\oplus t,w),
\eeqa
to obtain the proper solution of the master equation as
\beqa
S_{W_3} & =S_0 & +\int d^2x \int dz \{
 B^{\star\mu}[\del_\mu\eta_2 + e_\mu\del \eta_2
-2\del e_\mu \eta_2-\del B_\mu\eta_1 +2B_\mu\del \eta_1]    \nonumber \\
& & +e^{*\mu}[\del_\mu \eta_1 +e_\mu\del \eta_1 -\del e_\mu\eta_1
+2t(B_\mu\del \eta_2 -\del B_\mu \eta_2)
-\ep_{\mu\nu}e^{\star \nu}\eta_2\del \eta_2]     \nonumber \\
 & & +\eta^\star_1[\eta_1\del\eta_1 -2t\eta_2\del \eta_2]
+\eta^\star_2[\eta_1\del\eta_2 -2\del \eta_1 \eta_2]        \nonumber \\
& & -t^\star [2t\del\eta_1 +\del t \eta_1 +3w\del\eta_2
+2\del w\eta_2] \nonumber \\
& & -w^\star[\del w \eta_1 +3w\del\eta_1+2t\del t \eta_2 +2tt\del\eta_2]\}.
\eeqa

Gauge transformations of this theory are
in agreement with the ones given
in  \cite{SSN}. Of course, in \cite{SSN} the ghost fields
$\eta_1$, $\eta_2$ and $e^\star e^\star$ terms are absent.

To discuss gauge fixing conditions of this theory, we enlarge the
configuration space by introducing the fields
\beqa
\bar{\eta}_{a(0,-1)},\ \pi_{a(0,0)} & ; &
\ep (\eta_a)=1,\  \ep (\pi_a)  =0, \nonumber \\
\bar{\eta}_{a(2,0)}^\star ,\ \pi_{a(2,-1)}^\star &
; & \ep (\eta_a^\star )=0,\  \ep (\pi^\star_a)  =1, \nonumber
\eeqa
where $\ep$ denotes the Grassmann parity. Let us  define
\[
S_e=S+\int d^2 x \int dz {\bar{\eta}}^\star_a \pi_a,
\]
which is still a solution of the master equation.
Now, gauge fixing can be achieved in terms of the gauge fixing fermion
\[
\Psi_{(0,-1)} ,\ \ep (\Psi )=1,
\]
after substituting the antifields in $S_e$ by
\[
\chi^\star =\fr{\del \Psi (\chi )}{\del \chi },
\]
where $\chi$ represents the fields of the theory.

Of course, there exist different choices for gauge fixing fermion.
Let us first deal with
\be
\Psi =
{\bar{\eta}}_1 e_1  + {\bar{\eta}}_2 B_1.
\ee
In this gauge, the partition function after integrating over
$\pi_a,\ e_1,\ B_1,$ is
\be
Z=\int \prod dt dw de_odB_0
d\eta_a d {\bar{\eta}}_a
exp \int d^2x \int dz [
t\del_1e_0 +w\del_1B_0
+{\bar{\eta}}_a \del_1\eta_a ].
\ee
To calculate the path integral let us give the mode expansion of
the remaning fields:
\beqa
e_0(z)=\sum_{-\infty}^\infty e^n_0 z^{n+1} & ;  &
\eta_1(z)=\sum_{-\infty}^\infty \eta_1^n z^{n+1}  \nonumber \\
B_0(z)=\sum_{-\infty}^\infty B^n_0 z^{n+2} & ;  &
\eta_2(z)=\sum_{-\infty}^\infty \eta_2^n z^{n+2}  \nonumber \\
t  (z)=\sum_{-\infty}^\infty t^n   z^{-n-2} & ;  &
{\bar{\eta}}_1(z)=\sum_{-\infty}^\infty
{\bar{\eta}}^n_1 z^{-n-2}  \nonumber \\
w  (z)=\sum_{-\infty}^\infty w^n   z^{-n-3} & ;  &
{\bar{\eta}}_2(z)=\sum_{-\infty}^\infty
{\bar{\eta}}^n_2 z^{-n-3}.              \nonumber
\eeqa
Mode expansions of the ghosts $\eta_a$ and the antighosts
${\bar{\eta}}_a$, follow from the fact that
mode expansion of the generalized fields $\ti{h}$ and
$\ti{\Phi}$ are dictated by mode expansions of the
original fields $e,\ B,$ and $t,\ w$. Observe that in this
gauge ${\bar{\eta}}_a$ behave like $e^\star , B^\star$.

Hence, after performing the $z$ integral one obtains
\be
Z=\int \prod dt^n dw^n de_o^ndB_0^n
d\eta^n_ad{\bar{\eta}}^n_a
exp \int d^2x \sum_{-\infty}^\infty [
t^n\del_1e_0^n +w^n\del_1B_0^n
+{\bar{\eta}}^n_a \del_1\eta_a^n ].
\ee
Now, we can integrate over all of the remaining fields and find
\be
Z=\sum_{[G]}  {\rm sign}_G(\Delta ),
\ee
where
\[
\Delta = \lim_{N\rightarrow \infty}
\fr{{\rm det}^N\del_1{\rm det}^N\del_1}
{|{\rm det}^N\del_1{\rm det}^N\del_1| } .
\]
As a matter of fact,
the main point is to clarify what is meant by the
set $[G]$. This should be decided due to the properties of
the target manifold\cite{BBRT}. But this is out of
the scope of this work.

One can also choose the gauge fixing fermion as
\be
\Psi =
{\bar{\eta}}_1 e_0  + {\bar{\eta}}_2 B_0 ,
\ee
which leads to temporal gauge:
This choice is studied in \cite{III} on general grounds, and
it is shown that up to a field redefinition, effectively it yields a
free field theory.

Before studing the topological properties, it is not
possible to decide which gauge is the most interesting one.
Moreover, one should quantize the related BRST charge which
would follow after using (\ref{f})-(\ref{w3bl}) in
(\ref{omo1}). In the
case of finding an operator ordering yielding
a nilpotent BRST operator one should use it
to work out the physical states.
These subjects are left for future studies.

\vspace{2cm}

\begin{center}
{\bf Acknowledgment}
\end{center}

This work is partially supported by Turkish Scientific and Technological
Research Council (T\"{U}B\.{I}TAK) under TBAG/\c{C}G-1.

\pagebreak

\newcommand{\bi}{\bibitem}

\end{document}